# Length measurement of a moving rod by a single observer without assumptions concerning its magnitude

Bernhard Rothenstein and Ioan Damian
Politehnica University of Timisoara, Department of Physics, Timisoara, Romania

**Abstract.** *We extend the results presented by Weinstein[1] concerning the measurement of the length of a moving rod by a single observer, without making assumptions concerning the distance between the moving rod and the observer who measures its length.*

## 1. Who measures what?

Weinstein[1] considers a rod of proper length $L_0$ at rest in the K'(X'O'Y') reference frame. It is parallel to the O'X' axis, positioned apart from it at a distance $h$ (Figure 1). An observer $R'_0(0,0)$ located at the origin O' and at rest relative to K' measures its proper length $L_0$. Let $2'(x'_2 = r'_2 \cos\theta'_2, y'_2 = r'_2 \sin\theta'_2)$ be the leading edge of the rod and $1'(x'_1 = r'_1 \cos\theta'_1, y'_1 = r'_1 \sin\theta'_1)$ its trailing edge. Observer $R'_0(0,0)$ detects the light signals from both ends of the rod at a single time $t'_1 = 0$. Light originating from end 2' has left it at $t'_2 = -\dfrac{r'_2}{c}$ whereas light originating from 1' has left it at a time $t'_1 = -\dfrac{r'_1}{c}$. The events involved in the length measurement by $R'_0$ are $1'\left(r'_1\cos\theta'_1, r'_1\sin\theta'_1, -\dfrac{r'_1}{c}\right)$ and $2'\left(r'_2\cos\theta'_2, r'_2\sin\theta'_2, -\dfrac{r'_2}{c}\right)$. Consider the same experiment from the K(XOY) reference frame. The axes of the two reference frames are parallel to each other, the OX(O'X') axes are common, and K' moves with constant velocity $V$ relative to K in the positive direction of the common axes. What we compare are the length $L$ of the rod measured in K and its proper length $L_0$ measured in K'. The final results presented by Weinstein[1] are the consequence of some assumptions concerning how big or how small is the length of the position vector $r'_1$. The purpose of our paper is to present results free of such an assumption.

At the common origin of time in the two frames ($t=t'=0$) an observer $R_0(0,0)$ at rest in K and located at its origin O is located in front of observer $R'_0(0,0)$. From his point of view the events 1' and 2' defined above are characterized by the space-time coordinates $1(r_1\cos\theta_1, r_1\sin\theta_1, -\dfrac{r_1}{c})$ and $2(r_2\cos\theta_2, r_2\sin\theta_2, -\dfrac{r_2}{c})$. In accordance with the Lorentz-Einstein transformations we should have

$$x_1 = \gamma(x'_1 - \beta r'_1) \qquad (1)$$
$$x_2 = \gamma(x'_2 - \beta r'_2) \qquad (2)$$

with the usual relativistic notations $\beta = \dfrac{V}{c}$ and $\gamma = \dfrac{1}{\sqrt{1-\dfrac{V^2}{c^2}}}$. By definition

$$L_0 = (x_2' - x_1') \tag{3}$$

represents the proper length of the rod and

$$L = x_2 - x_1 \tag{4}$$

represents the measured length of the rod, following that measurement technique. From (3) and (4) we obtain

$$L = \gamma L_0 - \gamma \beta (r_2' - r_1') . \tag{5}$$

Application of the cosine theorem gives

$$r_2' = r_1' \sqrt{1 + \left(\dfrac{L_0}{r_1'}\right)^2 + 2\left(\dfrac{L_0}{r_1'}\right)\cos\theta_1'} . \tag{6}$$

Expressed as a function of $r_1'$ (6) becomes

$$\dfrac{L}{L_0} = \gamma \left\{ 1 - \beta \left(\dfrac{r_1'}{L_0}\right)\left[\sqrt{1 + \left(\dfrac{L_0}{r_1'}\right)^2 + 2\left(\dfrac{L_0}{r_1'}\right)\cos\theta_1'} - 1\right] \right\} \tag{7}$$

and we can express (7) as function of the angle $\theta_1$ measured in K via the aberration of light formula

$$\cos\theta_1' = \dfrac{\cos\theta_1 + \beta}{1 + \beta\cos\theta_1} . \tag{8}$$

The results become more transparent if we express (7) as a function of the distance h via

$$r_1' = \dfrac{h}{\sin\theta_1'} . \tag{9}$$

In the case of a rod standing "very close" to the observer ($\theta_1' = 0$, both ends of the rod are "outgoing") (7) leads to

$$\dfrac{L}{L} = \sqrt{\dfrac{1-\beta}{1+\beta}} . \tag{10}$$

In the case when both ends are "incoming" ($\theta_1' = \pi$) (7) leads to

$$\dfrac{L}{L} = \sqrt{\dfrac{1+\beta}{1-\beta}} \tag{11}$$

We can consider that $L_{0,inc} = \alpha L_0$, $\alpha < 1$, represents the length of the incoming part of the rod and that $L_{0,out} = (1-\alpha)L_0$ is the proper length of the outgoing part. The length of the rod measured under such conditions is

$$L = \gamma\alpha L_0(1-\beta) + \gamma(1-\alpha)L_0(1+\beta) = \gamma L_0\left[(1+\beta) - 2\alpha\beta\right] \tag{12}$$

Asking if observer $R_0(0,0)$ could detect the Lorentz contracted length of the rod

$$L_{L,c} = \sqrt{1 - \dfrac{V^2}{c^2}} L_0 \tag{13}$$

we impose that condition to (12) and the result is that for

$$\alpha = \frac{1+\beta}{2} < 1$$

the Lorentz contracted length of the rod is detected.

If the rod is located at a very large distance from the OX(O'X') axis ($\frac{L_0}{r_1'} \to 0$) we can consider that

$$\sqrt{1 + \left(\frac{L_0}{r_1'}\right)^2 + 2\left(\frac{L_0}{r_1'}\right)\cos\theta_1'} \cong 1 + \frac{L_0}{r_1'}\cos\theta_1'. \tag{14}$$

Under such conditions (7) becomes

$$\frac{L}{L_0} = \gamma(1 - \beta\cos\theta_1') = \frac{\gamma^{-1}}{1+\beta\cos\theta_1} \tag{15}$$

in accordance with Weinstein's[1] final result.

We present in Figure 2 the variation of $\frac{L}{L_0}$ with $\theta_1'$ for the same value of $\beta = 0.6$ and different values of $\frac{h}{L_0}$ showing the way in which $\frac{h}{L_0}$ influences the measurements. Intersecting the corresponding curves with $\frac{L}{L_0} = \sqrt{1-\beta^2}$, $\frac{L}{L_0} = 1$ and $\frac{L}{L_0} = \frac{1}{\sqrt{1-\beta^2}}$ we determine the values of $\theta_1'$ for which we detect Lorentz contraction, no change in length and Lorentz dilation respectively. For high values of $\frac{h}{L_0}$, the curves account for (15).

The basic reason that Lorentz-contraction and Lorentz-dilation appear in this case and not in the case discussed by Terrell[2] is that the length of the rod is here determined partly from angle measurements and partly from time of propagation of light measurement.

Sherwin[3] considers the presentation of the Lorentz contraction on a pulsed radar system. A source of light at rest in K' and located at its origin O' emits light signals in all directions of the X'O'Y' plane. Two of them detect the ends of the rod generally at different times and that enables a single observer to determine the radar measured length of the rod. In order to recover Sherwin's results we change the sign of $\beta$ in the equations derived above (the light signals are in this case "outgoing").

**2. Conclusions**

The reality of the Lorentz-contraction and Lorentz-dilation revealed above consists in the fact that they are the result of measurements and of the fact that Nature prevents us from results that would violate the sacrosanct principle of relativity. They convince us that when we speak about the magnitude of a physical quantity we should mention the observer (the observers) who measures it, when and where the measurements are performed and the measuring devices they use.

We consider that our paper is an introduction to a more elaborated approach presented recently by Deissler[4].

Figure 1.
A rod of proper length $L_0$ is at rest in the K'(X'O'Y') reference frame and located at a distance $h$ from the O'X' axis, parallel to it. An observer $R_0^{'}(0,0)$ located at the origin O' measures its length by receiving at $t'=0$ light signals that have left the edges of the rod at different times.

Figure 2.
The variation of the quotient $\frac{L}{L_0}$, between the measured length and the proper length, with the polar angle $\theta_1^{'}$ for $\beta = 0.6$ and different values of the parameter $\frac{h}{L_0}$.

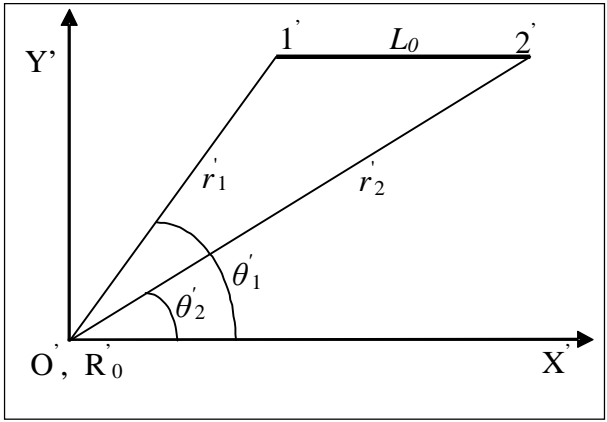

Fig. 1.

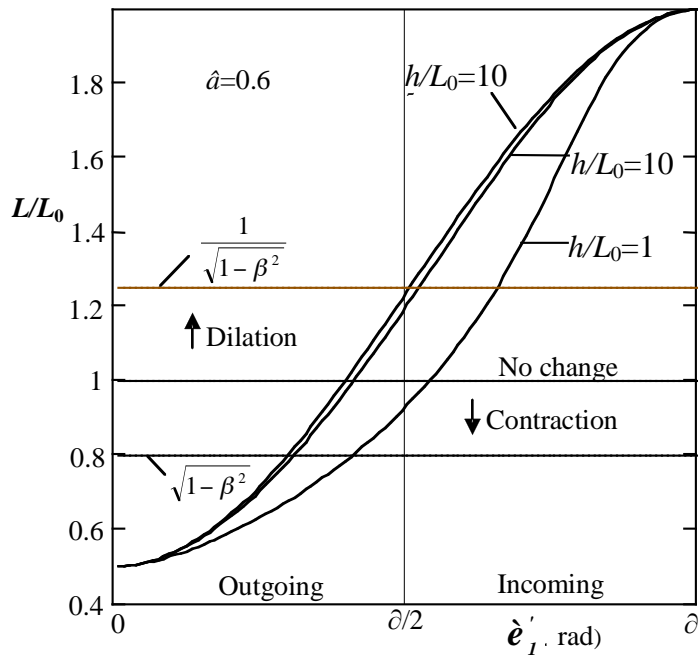

Fig.2.